\documentclass[11pt,preprint]{aastex}
\usepackage{epsfig}

\newcommand       \Angstrom     {\,{\rm \AA}}          

\newcommand       \cm           {\,{\rm cm}}
\newcommand       \mm           {\,{\rm mm}}

\newcommand	  \g		{\,{\rm g}}

\newcommand       \K            {\,{\rm K}}
\newcommand       \kms		{\,{\rm km\,s}^{-1}}

\newcommand	  \s		{\,{\rm s}}

\newcommand       \Qabs	        {Q_{\rm abs}}
\newcommand       \simlt        {\lesssim}
\newcommand       \simgt        {\gtrsim}

\newcommand	  \um	        {\mu{\rm m}}
\newcommand       \mum          {\,{\rm \mu m}}

\newcommand	  \mre		{m^{\prime}}
\newcommand	  \mim		{m^{\prime\prime}}
\newcommand	  \epsre	{\epsilon_1}
\newcommand	  \epsim	{\epsilon_2}

\newcommand	  \cabs		{C_{\rm abs}}
\newcommand	  \sigmaabs     {\sigma_{\rm abs}}
\newcommand	  \amin		{a_{\rm min}}
\newcommand	  \lmax		{l_{\rm max}}
\newcommand	  \omegap	{\omega_{\rm p}}
\newcommand	  \vF	        {v_{\rm F}}
\newcommand	  \nd	        {n_{\rm d}}
\newcommand	  \ndo	        {n_{\rm d}(0)}
\newcommand	  \zo	        {z_{\rm 0}}
\newcommand	  \To	        {T_{\rm 0}}
\newcommand	  \qo	        {q_{\rm 0}}
\newcommand	  \ho	        {H_{\rm 0}}
\newcommand	  \Mpc	        {{\rm Mpc}}
\newcommand	  \fdd	        {f_{\rm d}}
\newcommand	  \tauext	{\tau_{\rm ext}}
\newcommand	  \ltoa	        {\left(l/a\right)}
\newcommand	  \ltoamin	{\left(l/a\right)_{\rm min}}
\newcommand	  \ltoamax	{\left(l/a\right)_{\rm max}}
\newcommand	  \ltoalow	{\left(l/a\right)_{\rm low}}
\newcommand	  \ltoaupp	{\left(l/a\right)_{\rm upp}}
\newcommand	  \rhod	        {\rho_{\rm d}}
\newcommand	  \Omegad	{\Omega_{\rm d}}
\newcommand	  \Vtot	        {V^{\rm tot}}
\newcommand	  \lambdac	{\lambda_{\rm c}}
\newcommand       \etacut       {\eta_{\rm cut}}
\newcommand	  \xiqm	        {\xi_{\rm QM}}
\newcommand{\figwidth}{4.0in}

\countdef\decade=200
\advance\decade by \year
\countdef\hours=201
\advance\hours by \time
\divide\hours by 60
\countdef\mins=202
\advance\mins by \hours
\multiply\mins by 60
\multiply\hours by 100
\countdef\miltime=203
\advance\miltime by \hours
\advance\miltime by \time
\advance\miltime by -\mins

\shorttitle{Cosmic Needles}

\begin{document}

\title{
 \vspace*{-2.0em}
  {\normalsize\rm submitted to {\it The Astrophysical Journal Letters}}\\
  \vspace*{1.0em}
Cosmic Needles versus Cosmic Microwave Background Radiation
	 }

\author{Aigen Li}
\affil{Theoretical Astrophysics Program,
       University of Arizona, Tucson, AZ 85721\\
       {\sf agli@lpl.arizona.edu}}

\begin{abstract}
It has been suggested by a number of authors that the 2.7$\K$
cosmic microwave background (CMB) radiation might have arisen from the
radiation from Population III objects thermalized by conducting
cosmic graphite/iron needle-shaped dust. Due to lack of an accurate 
solution to the absorption properties of exceedingly elongated grains,  
in existing literature which studies the CMB thermalizing process 
they are generally modelled as (1) needle-like spheroids in 
terms of the Rayleigh approximation; (2) infinite cylinders;
and (3) the antenna theory.
We show here that the Rayleigh approximation is not valid since 
the Rayleigh criterion is not satisfied for highly conducting needles.
We also show that the available intergalactic iron dust,
if modelled as infinite cylinders, is not sufficient to 
supply the required opacity at long wavelengths to obtain
the observed isotropy and Planckian nature of the CMB. 
If appealing to the antenna theory, conducting iron needles with 
exceedingly large elongations ($>10^4$) appear able to provide
sufficient opacity to thermalize the CMB within the iron density 
limit. But the applicability of the antenna theory  
to exceedingly thin needles of nanometer/micrometer 
in thickness needs to be justified.
\end{abstract}


\section{Introduction: Absorption Properties of Cosmic 
Needles\label{sec:intro}}
The 2.7$\K$ cosmic microwave background (CMB) is generally 
interpreted as being relic radiation from the early hot universe
of a big bang origin. Alternative attempts at explaining the 
observed CMB as a post-big bang phenomenon have been continuously 
made in terms of emission from ``Population III'' objects at high 
redshift (presumably either a pre-galactic generation of very 
massive stars or black hole accretion flows) 
thermalized by hollow spheres (Layzer \& Hively 1973) 
or long slender conducting cosmic whiskers or 
``cosmic needles'' (Hoyle, Wickramasinghe, \& Reddish 1968; 
Wickramasinghe et al.\ 1975; Rana 1980; Wright 1982; 
Hoyle, Narlikar, \& Wickramasinghe 1984;
Hawkins \& Wright 1988; Hoyle \& Wickramasinghe 1988; 
Bond, Carr, \& Hogan 1991; Wickramasinghe 1992; Wickramasinghe et al.\ 1992; 
Wickramasinghe \& Hoyle 1994; Aguirre 2000).

The reason for invoking ``conducting needles'' is because neither spherical 
grains (both dielectric and metallic) nor dielectric needles have
high enough opacity in the far infrared (IR) to be an efficient 
thermalizing agent unless an unreasonably large amount
of dust is invoked. This can be seen from the absorption cross section 
expressions of spherical or spheroidal grains.
Let $\epsilon(\lambda) = \epsre + i\epsim$ be the dust complex 
dielectric function at wavelength $\lambda$. 
The absorption cross section $\cabs$ per unit volume ($V$) for
spheres in the Rayleigh regime (Bohren \& Huffman 1983) is
\begin{equation}
\cabs/V \approx \frac{18\pi}{\lambda}\frac{\epsim}{(\epsre+2)^2+\epsim^2} ~~~.
\end{equation}
For dielectric spheres, $\cabs \propto \lambda^{-1}\epsim \propto 
\lambda^{-2}$ approaches zero as $\lambda\rightarrow \infty$ 
since at far-IR $\epsre$ approaches a constant
$\gg\epsim$ while $\epsim \propto \lambda^{-1}$;
for metallic spheres with a conductivity of $\sigma$, 
$\cabs\propto \lambda^{-1}\epsim^{-1}\propto \lambda^{-2}$ 
also approaches zero as $\lambda\rightarrow \infty$ 
since $\epsim = 2\lambda\sigma/c \propto \lambda$ 
($c$ is the speed of light) and $\epsre \ll \epsim$.
Let needle-shaped grains be represented by thin prolate
spheroids of semiaxes $l$ along the symmetry axis
and $a$ perpendicular to the symmetry axis. 
In the Rayleigh limit, the absorption cross section 
per unit volume for needle-like prolate grains ($l\gg a$) is
approximately
\begin{equation}\label{eq:needle}
\cabs/V \approx \frac{2\pi}{3\lambda}\frac{\epsim}
{\left[L_{\|}(\epsre-1) + 1\right]^2 
+ \left(L_{\|}\epsim\right)^2}
\end{equation} 
where $L_{\|}\approx \left(a/l\right)^2\ln(l/a)$ is the
depolarization factor parallels to the symmetry axis.
For dielectric needles, $\cabs \propto \lambda^{-1}\epsim
\propto \lambda^{-2}$ at far-IR since we usually have
$L_{\|}(\epsre-1) + 1 \gg L_{\|}\epsim$ while $\epsre$ is
insensitive to $\lambda$ at long wavelengths;
for metallic needles, it appears at first glance that, for a given
value of $\epsim$ (at a given $\lambda$) -- no matter how large -- 
one can always find a sufficiently long needle with 
$L_{\|}\epsim \simlt 1$ and
$L_{\|}(\epsre-1) \ll 1$ (Greenberg 1972) so that
$\cabs \propto \lambda^{-1}\epsim \propto \sigma$
which can be very large.\footnote{%
 This should not be considered inconsistent with the 
 Kramers-Kronig relation since for a given elongation $\ltoa$ 
 there exists a long-wavelength cutoff ($\lambda_0$) for $\cabs$: 
 $\cabs \propto \lambda^{-2}$ as 
 $\lambda > \lambda_0$ (see Eqs.[\ref{eq:wright}-\ref{eq:cabs0}]).
 }
Therefore, it is possible for metallic needles with
high electrical conductivities to provide a large quantity of
opacity at long wavelengths to thermalize the cosmic background.

For conducting needles Eq.(\ref{eq:needle})
can be expressed as (Wright 1982) 
\begin{equation}\label{eq:wright}
\cabs = \frac{\cabs^{0}}{1 + \left(\lambda/\lambda_0\right)^2}
\end{equation} 
where the long-wavelength cutoff $\lambda_0$ is
\begin{equation}\label{eq:cutoff}
\lambda_0 \equiv \frac{\rho c}{2}
          \frac{1 + L_{\|}\left(\epsre-1\right)}{L_{\|}} 
          \approx \frac{\rho c}{2} 
          \frac{\left(l/a\right)^2}{\ln\left(l/a\right)} 
\end{equation}
and
\begin{equation}\label{eq:cabs0}
\cabs^0 \equiv \frac{4\pi V}{3\rho c}
        \frac{1}{\left[1 + L_{\|}\left(\epsre-1\right)\right]^2}
        \approx \frac{4\pi V}{3\rho c}
\end{equation}
where $\rho = 1/\sigma$ is the dust material resistivity.
It is seen that Eq.(\ref{eq:cutoff}) establishes a lower bound 
on the elongation $l/a$ of the needles which absorb strongly at
wavelengths out to $\lambda_0$.

Eqs.(\ref{eq:needle}-\ref{eq:cabs0}) have been widely used in 
obtaining the dust opacity in the far-IR and microwave range
(Rana 1980; Wright 1982; Hawkins \& Wright 1988; 
Hoyle \& Wickramasinghe 1988; Bond, Carr, \& Hogan 1991; 
Wickramasinghe et al.\ 1992; Wickramasinghe \& Hoyle 1994). 
However, none of these has explicitly taken 
into account the criterion to which the Rayleigh approximation 
is applicable (Bohren \& Huffman 1983):
\begin{equation}\label{eq:rayleigh}
\frac{2\pi l}{\lambda} \ll 1 ~; ~~~ |m|\frac{2\pi l}{\lambda} \ll 1 
\end{equation}
where $m(\lambda) = \mre + i\mim$ is the complex 
refractive index ($\epsilon = m^2$). For metals at long wavelengths
we have $\mre \approx \mim \approx \left(\sigma\lambda/c\right)^{1/2}$.
Therefore, the Rayleigh approximation (Eq.[\ref{eq:rayleigh}])
establishes an upper bound on the needle length:
\begin{equation}\label{eq:lmax}
l \ll \frac{1}{2\pi}\left(\frac{\lambda c}{\sigma}\right)^{1/2}
      = \frac{1}{2\pi}\left(\lambda \rho c\right)^{1/2} ~~~.
\end{equation}
The reason for applying this criterion for limiting the needle size 
is that, when it is not satisfied, the cross sections given by 
Eq.(\ref{eq:needle}) are overestimates of the true cross sections. 
It is only when all elements within the particle radiate in phase 
with each other (i.e. negligible phase shift of light within the
particle) that we can achieve the high absorptivities (Greenberg 1980). 
The implication of Eq.(\ref{eq:lmax}) for cosmic needles is significant. 
For example, for iron needles of 
$\rho = 10^{-16}\s$ to absorb efficiently out to
$\lambda_0 = 5\mm$, Eq.(\ref{eq:cutoff}) requires an elongation
of $l/a \approx 1600$ (also see Wright 1982). To satisfy the
Rayleigh criterion, Eq.(\ref{eq:lmax}) leads to $l \ll 1.9\mum$.
A combination of Eq.(\ref{eq:cutoff}) and Eq.(\ref{eq:lmax})
requires the needle radius $a \ll 12\Angstrom$. It is unlikely 
that such tiny iron needles exist in astrophysical environments.
After all, for stacks of layers of $2\times 2$ and $3\times 3$ iron 
atoms the needle radius would already be $\approx 2.8, 4.2\Angstrom$, 
respectively. To be conservative, we therefore take the minimum radius 
of iron needles to be $\amin = 3.5\Angstrom$. 
In the following text, we will take the Rayleigh criterion to be 
\begin{equation}\label{eq:lmax0}
|m|\frac{2\pi \lmax}{\lambda} \approx 0.1 ~; ~~~
\lmax \approx \frac{1}{20\pi}\left(\lambda \rho c\right)^{1/2}
\end{equation}
where $\lmax$ is the maximum value of the needle length $l$
for which the Rayleigh approximation is still valid. 

\begin{figure}[h] 
\begin{center}
\epsfig{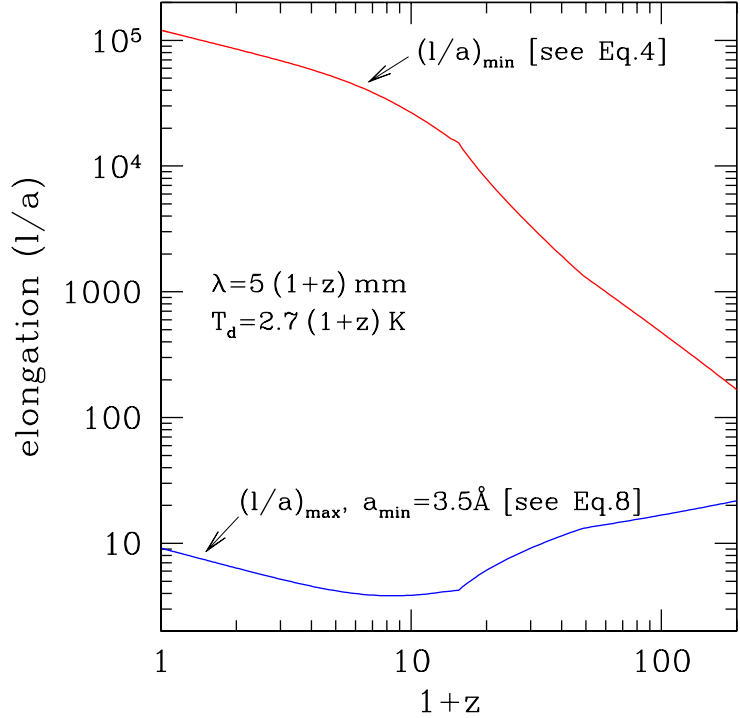}
\end{center}\vspace*{-1em}
\caption{
        \label{fig:l2aminmax}
        \footnotesize
        Lower [$\ltoamin$] and upper [$\ltoamax$] limits on 
        the elongation of iron needles (assumed to thermalize 
        background radiation emitted at $z$ and observed at 
        $\lambda=5\mm$)         
        as a function of redshift $z$ obtained respectively 
        from the long-wavelength opacity consideration 
        (Eq.[\ref{eq:cutoff}]) and the Rayleigh criterion 
        (Eq.[\ref{eq:lmax0}]). 
        It is apparent that the Rayleigh 
        approximation (Eq.[\ref{eq:needle}]) is not valid for
        studies of the CMB thermalization by iron needles
        since $\ltoamax \ll \ltoamin$.
        }
\end{figure}

For a given wavelength $\lambda$ and a given dust conductivity
$\sigma$ which is dependent on dust material and temperature
we can obtain from Eq.(\ref{eq:cutoff}) $\ltoamin$ -- the lower 
limit on the needle elongation which displays appreciable opacity
at wavelengths up to $\lambda$ (following Wright 1982, we take 
$\lambda=5\mm$ for discussion); and from Eq.(\ref{eq:lmax0})
$\ltoamax = \lmax/\amin$ -- the upper limit on the needle elongation
to which the Rayleigh approximation (Eq.[\ref{eq:needle}]) 
is applicable.  In Figure \ref{fig:l2aminmax} we present 
$\ltoamin$ and $\ltoamax$ for cosmic iron needles 
(see \S\ref{sec:ironnk}) thermalizing the background radiation 
emitted at redshift $z$ and observed at wavelength $\lambda=5\mm$.
It is seen in Figure \ref{fig:l2aminmax} that even with
$\amin=3.5\Angstrom$, $\ltoamax \ll \ltoamin$ for $z$ up to 200.
This clearly indicates that it is not appropriate to adopt the Rayleigh 
approximation (Eq.[\ref{eq:needle}]) when studying the CMB thermalization 
by cosmic iron needles.


The absorption cross sections of needles may be
approximated by those of infinite cylinders provided $l$
exceeds $a$ by a factor of $\sim 4$ (Wickramasinghe 1973)
or $\sim 9$ (Lind \& Greenberg 1966). 
For needles with a radius less than $\sim 10\Angstrom$ 
the classical scattering theory does not apply (Platt 1956; 
Greenberg 1960). We will adopt the infinite cylinder results 
but take a cutoff at $\lambdac \approx 400\,l$ (Platt 1956).
Similar to Li \& Draine (2001), we assume a continuous distribution 
for the absorption properties of classic infinite cylinders and 
Platt particles 
($a\simlt 10\Angstrom$):
\begin{equation}\label{eq:cabsnew}
\cabs(\lambda) = \cabs^{\rm inf}\left[\xiqm\etacut
                  + \left(1-\xiqm\right)\right] ~,
\end{equation}
\begin{equation}\label{eq:xiqm}
\xiqm(a) = {\rm min}\left[1, (a_{\xi}/a)^3\right],
           ~~~~ a_{\xi} = 10{\rm \AA} ~,
\end{equation}
\begin{equation}\label{eq:etacut}
\etacut(\lambda, \lambdac) = \frac{1}{\pi} 
\arctan\left[\frac{10^3 (y-1)^3} {y}\right] + \frac{1}{2}, ~~~ 
y=\lambdac/\lambda, \lambdac = 400\,l
\end{equation}
where $\cabs^{\rm inf} (=2a \Qabs)$ is the absorption cross sections for
infinite cylinders ($\Qabs$ is the absorption cross section per unit
length divided by $2a$), and $\etacut$ is a cutoff function. 

In the far-IR and microwave regions, for metallic needles
both inductance and charge separation effects are ignorable
(Hoyle \& Wickramasinghe 1988).
Therefore, conducting needles can be treated as an antenna 
(Wright 1982):
$\cabs/V = 4\pi/\left(3\rho c\right)$ for $\lambda < \lambda_0$
where $\lambda_0$ is the same as in Eq.(\ref{eq:cutoff}).
Taking into account the quantum mechanical effect,
the absorption cross section of an antenna can be expressed as
\begin{equation}\label{eq:ant}
\cabs(\lambda)/V = \frac{4\pi}{3\rho c}\left[\xiqm\etacut(\lambda,\lambdac)
                  + \left(1-\xiqm\right)\right]
                  /\left[1 + \left(\lambda/\lambda_0\right)^2\right]
\end{equation} 
where the $1/\left[1 + \left(\lambda/\lambda_0\right)^2\right]$ term 
accounts for the cutoff at $\lambda_0$ which has been
justified by Wright (1987).

\begin{figure}[h] 
\begin{center}
\epsfig{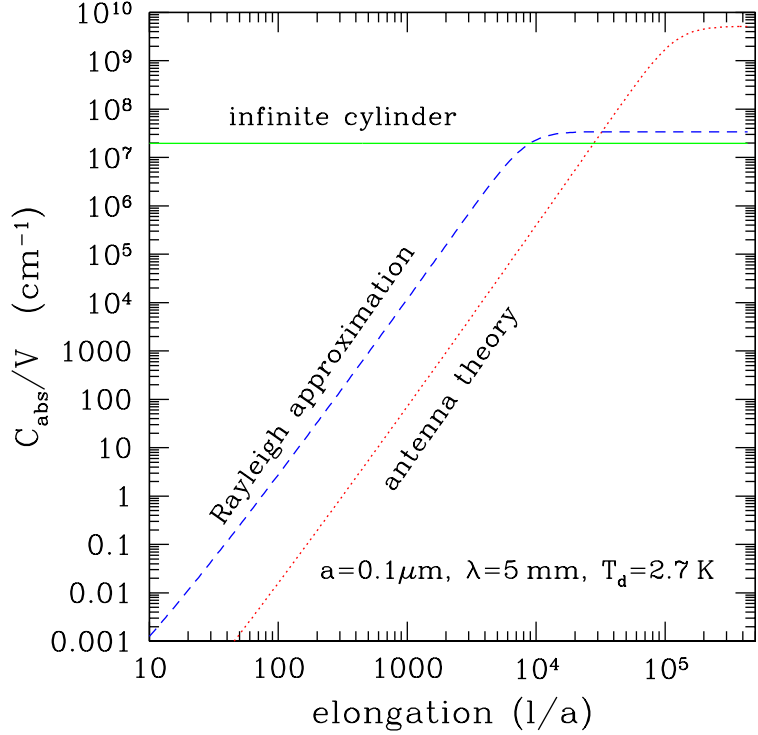}
\end{center}\vspace*{-1em}
\caption{
        \label{fig:cabs2v}
        \footnotesize
        Absorption cross sections per unit volume at $\lambda=5\mm$
        as a function of elongation for iron needles with a radius of 
        $a=0.1\mum$ at $T_{\rm d}=2.7\K$ calculated from
        the infinite cylinder representation (solid line), 
        Rayleigh approximation (dashed line), and antenna theory
        (dotted line). Note that the Rayleigh approximation results 
        should not be considered too seriously since the Rayleigh 
        criterion which requires $\ltoa < 0.03$ is never satisfied.
        }
\end{figure}

In Figure \ref{fig:cabs2v} we compare the absorption cross sections
(per unit volume) at $\lambda=5\mm$ calculated from infinite iron 
cylinders and iron antennae with a radius of $a=0.1\mum$ and a range 
of elongations at $T_{\rm d}=2.7\K$. Although the Rayleigh approximation
is not valid for iron needles capable of efficiently supplying far-IR
and microwave opacity, we also present in Figure \ref{fig:cabs2v} results
obtained from the Rayleigh approximation (Eq.[\ref{eq:needle}]) since
it is widely used in literature. It is seen in Figure \ref{fig:cabs2v}
that the infinite cylinder model predicts a much larger 5$\mm$ opacity 
for $\ltoa <3\times 10^4$ and a much smaller one for $\ltoa > 3\times 10^4$.
The antenna model shows a rapid drop at $\ltoa < 1.2\times 10^5$ 
which corresponds to a cutoff wavelength $\lambda_0\approx 5\mm$
(see Eq.[\ref{eq:cutoff}]). 
This is because the long-wavelength cutoff becomes smaller as $\ltoa$ 
decreases so that the iron opacity at 5$\mm$ decreases rapidly, too.  
This trend is also seen in the Rayleigh curve.\footnote{%
 The reason why the onset $\l/a$ value of the drop in the $\cabs/V$ curve
 for the Rayleigh approximation model differs from that for the 
 antenna model is because the second term in the r.h.s of 
 Eqs.(\ref{eq:cutoff}-\ref{eq:cabs0}) does not always hold.
 }
The dramatic differences among the absorption  cross sections
calculated from the three methods will have dramatic effects on
the CMB thermalization model (see \S\ref{sec:cmb}).

It is the purpose of this {\it Letter} first to show (see above) 
that the widely adopted Rayleigh approximation is not applicable 
to studies of the CMB as a result of thermalization by cosmic needles,
and then to estimate the quantity of dust required to thermalize the
background radiation using the absorption cross sections of 
either infinite cylinders or antennae.
We will only consider iron grains since they absorb more
efficiently in the far-IR than graphite and also because
the upper bound on the total amount of microwave radiation 
generated by graphite needles was shown considerably smaller than 
the observed CMB and it was also shown that the optical depth of 
the graphite needle-containing cloud is not sufficiently large 
for the cloud to radiate like a black body (Sivaram \& Shah 1985).
Condensed in supernova ejecta, iron grains may likely form 
as slender whiskers by the ``screw dislocation'' mechanism
which has been attested experimentally and (at least some fraction 
of them) are then expelled into extragalactic space without
significant destruction due to sputtering (see Hoyle \& Wickramasinghe
1988 and references therein).   
It is interesting to note that small iron particles were among 
the materials initially proposed to be responsible for 
the interstellar reddening, based on an analogy with small
meteors or micrometeorites supposedly fragmented into 
finer dust (Schal\'{e}n 1936; Greenstein 1938).
In \S\ref{sec:ironnk} we calculate the dielectric functions
of iron grains based on the Drude theory.
In \S\ref{sec:cmb} we calculate the extinction optical depth 
caused by cosmologically distributed iron needles and estimate 
the amount of iron needles required to thermalize the CMB and
compare it with the available intergalactic iron density.  
Concluding remarks are given in \S\ref{sec:conclusion}.

\section{Optical Properties of Iron Grains}\label{sec:ironnk}
We use the Drude free-electron model (Bohren \& Huffman 1983) to calculate 
the iron dielectric functions:
\begin{equation}\label{eq:drude}
\epsilon(\omega) = 1 - \frac{\left(\omegap\tau\right)^2}
{\left(\omega\tau\right)^2+i\omega\tau}
\end{equation} 
where $\omega=2\pi c/\lambda$ is the angular frequency;
$\omegap$ is the plasma frequency
\begin{equation}\label{eq:omegap}
\omegap^2 = \frac{4\pi e^2 n_e}{m_{\rm eff}}
\end{equation} 
where $e$ is the proton charge; $n_e$ is the density of free electrons;
$m_{\rm eff}$ is the effective mass of a free electron;
$\tau$ is the collision time 
\begin{equation}\label{eq:taucoll}
\tau^{-1} \approx \frac{\omegap^2}{4\pi \sigma} + \frac{\vF}{a}
\end{equation} 
where $\vF$ is the velocity of electrons at the Fermi surface
which we take to be $\vF = 10^8\cm\s^{-1}$.
The first term in the r.h.s of Eq.(\ref{eq:taucoll}) is for
bulk material; the second term accounts for the small size effect:
the inverse of the collision time is increased because of additional
collisions with the needle boundary. 

Lack of knowledge of the temperature dependence of $\omegap$,
we adopt the room temperature value of
$\omegap = 5.56\times 10^{15}\s^{-1}$ (Ordal et al.\ 1988).
We fit the temperature dependent electrical resistance $\rho(T)$
(Lide \& Frederikse 1994) by polynomials:
\begin{equation}\label{eq:high2c}
\left[\frac{\rho(T)}{10^{-18}\s}\right] =
\left\{\begin{array}{lr}
0.0286 - 7.57\ 10^{-4}\ T + 5.48\ 10^{-5}\ T^2 ~, &0\K < T < 40\K ~,\\
0.0759 - 8.96\ 10^{-3}\ T + 2.22\ 10^{-4}\ T^2 ~, & 40\K < T < 130\K ~,\\
-0.300 + 0.0120\ T + 8.38\ 10^{-5}\ T^2 ~, & 130\K < T < 1000\K ~.\\
\end{array}\right.
\end{equation}

\begin{figure}[h] 
\begin{center}
\epsfig{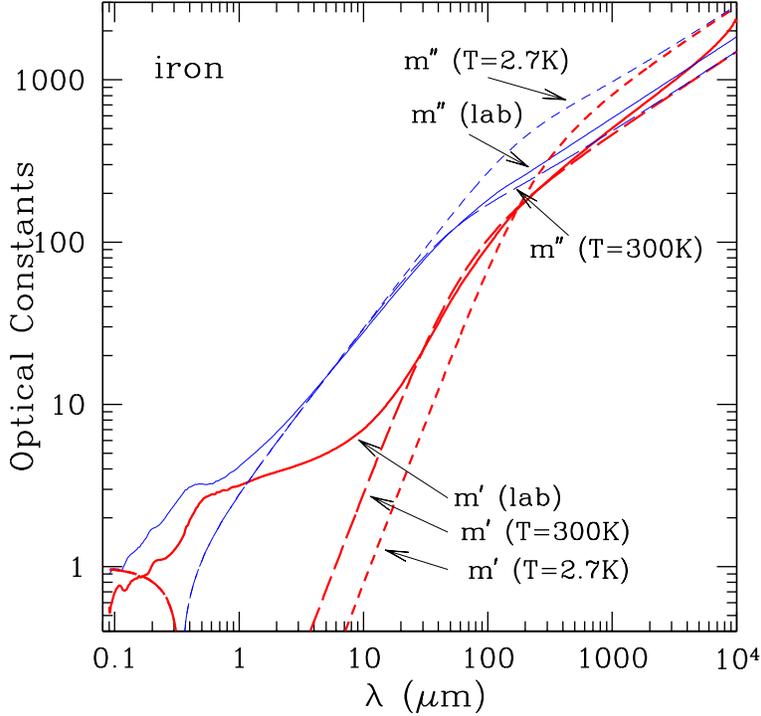}
\end{center}\vspace*{-1em}
\caption{
        \label{fig:ironnk}
        \footnotesize
        Optical constants $\mre$ (thick lines), $\mim$ (thin lines)
        of iron calculated for $T=2.7\K$ (dashed line) and
        $T=300\K$ (long dashed line) from the Drude free-electron model.
        Also shown is the experimental data (solid line) measured
        at room temperature. 
        }
\end{figure}

In Figure \ref{fig:ironnk} we show the refractive indices calculated
for $T=2.7\K, 300\K$ for $a=0.1\mum$. We have also taken the following 
``synthetic'' approach to obtain the room temperature complex refractive
index $m(\lambda)=\mre+i\mim$ for iron:
for $0.01 < \lambda < 0.248\mum$
we take $\mim$ of Moravec, Rife, \& Dexter (1976);
for $0.367 < \lambda < 0.667\mum$
we take $\mim$ of Gray (1972);
for $0.667 < \lambda < 285\mum$
we take $\mim$ of Ordal et al.\ (1988);
for $\lambda > 285\mum$ we approximate
$\mim(\lambda) \approx \mim(285\um)\left(\lambda/285\mum\right)^{1/2}$
since in the far-IR for metals
we have $\mim \propto \left(\lambda/\rho\right)^{1/2}$.
After smoothly joining the adopted $\mim$ (in so doing,
the Gray [1972] data is reduced by a factor of 1.2),
we calculate $\mre$ from $\mim$
through the Kramers-Kronig relation (Bohren \& Huffman 1983).
The results are also presented in Figure \ref{fig:ironnk}.
It is seen in Figure \ref{fig:ironnk} that the Drude free-electron model
provides a good approximation for $\lambda >20\mum$. 

\section{On Cosmic Needles Origin of CMB}\label{sec:cmb}
Following Draine \& Shapiro (1989), we define 
$\nd(z)\equiv \fdd(z)\ndo (1+z)^3$ as the number density of 
needle-shaped grains at redshift $z$, where $\ndo$ is a constant
and $\fdd$ is the ratio of the actual number density of grains
at $z$ to the number density if the grain number per comoving
volume were constant. The extinction optical depth for radiation
emitted at redshift $\zo$ and observed at wavelength $\lambda$ is
\begin{equation}\label{eq:opdpth}
\tauext(\zo,\lambda) = \left(\frac{c}{\ho}\right)\ndo
\int_{0}^{\zo} \sigmaabs\left(\frac{\lambda}{1+z},\To\left[1+z\right]\right)
\frac{\fdd(z)\left(1+z\right)dz}{\left(1+2\qo z\right)^{1/2}}
\end{equation}  
where $\ho$ is the Hubble constant, $\qo$ is the cosmological
deceleration parameter, $\To$ ($\approx 2.7\K$) is the current 
temperature of the microwave background, and $\sigmaabs$ is the 
absorption cross section\footnote{%
  Since the iron dielectric functions are dependent on temperature
  (see \S\ref{sec:ironnk}), the absorption cross sections are thus
  a function of temperature, too. We set $T=\To(1+z)$ because we want
  to thermalize the background, not to distort it (see Wright 1982).
  If iron needles act as a thermalizer to produce the $\To(1+z)$ 
  background, it is necessary that they are themselves thermalized
  to $\To(1+z)$.  
  }
integrated over a distribution of needle elongations $\ltoa$
\begin{equation}\label{eq:sigmaabs}
\sigmaabs(\lambda, T) = \int_{\ltoalow}^{\ltoaupp} 
\cabs(\lambda, T) \frac{dn\ltoa}{d\ltoa}d\ltoa
\end{equation} 
where the lower cutoff $\ltoalow$ is taken to be 10;
the upper cutoff $\ltoaupp$ is set at $10^5$ since laboratory-grown
iron needles display values of $\ltoa$ up to $\sim 10^5$
(see Hoyle \& Wickramasinghe 1988 and references therein; Agurrie 2000).
We assume a power-law distribution for the needle elongation
for $\ltoalow\leq \ltoa \leq \ltoaupp$:
\begin{eqnarray}\label{eq:dnde}
\frac{dn\ltoa}{d\ltoa} &=&
\frac{1}{\ln\left[\ltoaupp/\ltoalow\right]}\ltoa^{-1} ~, ~~\beta= 1 ~;
\\ 
&=& \frac{(1-\beta)}
{\ltoaupp^{1-\beta}-\ltoalow^{1-\beta}}\ltoa^{-\beta} ~, ~~\beta\neq 1 ~;
\end{eqnarray}
where $\beta$ is the power-law index. We will consider
two cases: $\beta =1$ (mass-equipartition distribution) 
and $\beta=3.5$; the former may arise from a plausible 
whisker formation process and can be maintained through 
an ongoing fragmentation process
(Wickramasinghe \& Wallis 1996; Aguirre 1999); the latter
may result from shattering from grain-grain collisions.

\begin{figure}[h] 
\begin{center}
\epsfig{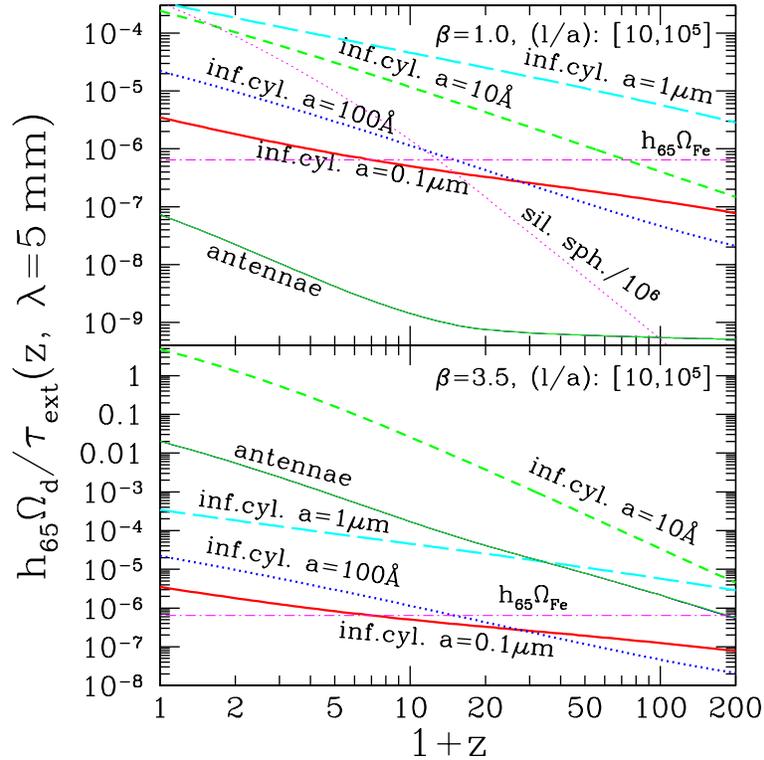}
\end{center}\vspace*{-1em}
\caption{
        \label{fig:omegad}
        \footnotesize
        The quantity ($h_{65}\Omega_{\rm d}$) of cosmic iron 
        needles as a function of $1+z$ required to produce 
        an optical depth $\tauext$ at $\lambda=5\mm$. 
        The absorption cross sections are calculated from 
        infinite cylinders with a radius of $a=0.1\mum$
        (heavy solid line), $a=100\Angstrom$ (heavy dotted 
        line), $a=10\Angstrom$ (heavy dashed line),
        and $a=1\mum$ (heavy long-dashed line), 
        and the antenna theory (thin solid line). We assume 
        a power-law distribution $dn/d(l/a) \propto (l/a)^{-\beta}$
        for the needle elongation with $\beta=1$ (upper panel)
        and $\beta=3.5$ (lower panel) with a lower cutoff at
        $\ltoalow = 10$ and a upper cutoff at $\ltoaupp=10^5$.
        For comparison, results for spherical ``astronomical 
        silicates'' (Draine \& Lee 1984) are also shown
        (reduced by a factor of $10^6$; thin dotted line). 
        The horizontal dot-dashed
        line shows the available intergalactic iron density
        $h_{65}\Omega_{\rm Fe}$. It is seen that the thermalization 
        condition $\tauext \gg 1$ at microwave wavelengths 
        required by the isotropy and Planckian nature of the CMB 
        is only attainable by the antenna theory for $\beta \simlt 2.5$. 
        However, we note that it has not been justified whether 
        the antenna theory is valid for extremely thin needles.        
        }
\end{figure}

Let $\rhod$ ($\approx 7.86\g\cm^{-3}$) be the mass density of 
iron grains, and $\Omegad$ be the 
ratio of the space-averaged comoving mass density of needles to 
the present critical density: $\Omegad\equiv \fdd(z)\ndo \Vtot\rhod
\left[8\pi G/3\ho^2\right]$ ($G$ is the Gravitation constant) where
\begin{equation}\label{eq:vtot}
\Vtot = \int_{\ltoalow}^{\ltoaupp} 
\psi\pi a^3 \ltoa\frac{dn\ltoa}{d\ltoa}d\ltoa
\end{equation}
where $\psi=1$ for cylinders and $\psi=4/3$ for prolates.
For simplicity, we take $\fdd(z)=1$ for $0 < z <\zo$ and 
$\fdd(z)=0$ otherwise (i.e. assuming a uniform comoving number 
of dust grains back to redshift $\zo$; see Draine \& Shapiro 1989).
Therefore, we can estimate the amount of iron needles required
to produce an optical depth of $\tauext$ at redshift $\zo$
and wavelength $\lambda$
\begin{equation}\label{eq:omega}
\Omegad = \left(\frac{8\pi G \rhod}{3c \ho}\right) 
\tauext(\zo,\lambda)/\int_{0}^{\zo} 
\sigmaabs\left(\frac{\lambda}{1+z},\To\left[1+z\right]\right)/\Vtot
\frac{\left(1+z\right)dz}{\left(1+2\qo z\right)^{1/2}} ~~~.
\end{equation}  
Adopting $\ho = 65\kms\Mpc^{-1}$ and $\qo=0.5$, we present in
Figure \ref{fig:omegad} the results on $h_{65}\Omegad/\tauext$ given 
by Eq.(\ref{eq:omega}) for $\lambda = 5\mm$ as a function of $z$
where $h_{65}\equiv \ho/100\kms\Mpc^{-1}=0.65$.
The absorption cross sections are obtained from infinite cylinders
(Eq.[\ref{eq:cabsnew}]) and antennae (Eq.[\ref{eq:ant}]) with a radius
of $a=10,100,1000, 10^4\Angstrom$. For comparison, results for spherical
``astronomical silicates'' (Draine \& Lee 1984) are also plotted.
It can be seen in Figure \ref{fig:omegad} that
(1) $\Omegad$ decreases with the increasing of $z$ as expected from
Eq.(\ref{eq:omega}); (2) for a given $z$, $\Omegad$ decreases with 
radius $a$ ($a\simlt 0.1\mum$) for infinite cylinders while it is 
insensitive to $a$ for antennae provided $a\simgt 10\Angstrom$ 
as expected from Eq.(\ref{eq:ant}); 
(3) for infinite cylinders with $a\simgt 10\Angstrom$ 
$\Omegad$ is insensitive to $\beta$ as expected from the way we calculate 
their absorption cross sections; for antennae $\Omegad$ increases with
$\beta$ since a higher $\beta$ implies fewer highly-elongated needles
and thus a lower opacity at $\lambda=5\mm$ since the long-wavelength cutoff
shifts to shorter wavelengths for needles with smaller elongations.

Aguirre (2000) argued that the density of intergalactic dust is
$\Omega_{\rm dust}\simlt 10^{-5}$. Assuming a solar-like metallicity
ratio pattern, this leads to an iron density of 
$\Omega_{\rm Fe}\simlt 10^{-6}$. In order to obtain the observed 
high degree of spacial isotropy and Planckian nature of the CMB, 
one requires the optical depth $\tauext \gg 1$ in the microwave 
region (Wickramasinghe et al.\ 1975; Rana 1980; Sivaram \& Shah 1985). 
It is clear from Figure \ref{fig:omegad} that it is unlikely for 
the intergalactic iron needles modelled as infinite cylinders to
produce $\tauext(\lambda=5\mm)\gg 1$ even if all available intergalactic 
iron have been locked up in needles since $\Omega_{\rm Fe}/\Omegad <1$
for $z<10-20$ and  $\Omega_{\rm Fe}/\Omegad$ is just about 10 
for $z\sim 100-200$.
It is also clear from the spherical silicate model that
direct thermalization of pregalactic starlight by 
intergalactic dielectric dust spherical in shape is not viable
since $\Omega_{\rm Si}/\Omegad \ll1$.

Wickramasinghe et al.\ (1992) argued that the infinite cylinder
approximation for slender needles is only valid when the inductance
and charge separation effects are not important which requires
$\lambda < \left(\rho c/4\right) \ltoa^2/\ln\ltoa$.
This would strengthen the conclusion drawn above for the infinite
cylinder model: there is simply not enough iron dust to produce a large 
optical depth at microwave range since with this taken into account
the absorption cross section (per unit dust mass) at this wavelength
region would be reduced. Increasing $\ltoalow$ and/or $\ltoaupp$
would have little effects on the infinite cylinder model.
Increasing the cylinder thickness ($a>0.1\mum$) either does not help
(see Figure \ref{fig:omegad}).

However, if appealing to the antenna theory, it appears that 
$\tauext(\lambda=5\mm)\gg 1$ is attainable by models with
a relatively flat distribution of needle elongation 
($\beta \simlt 2.5$) since the absorption cross sections per unit dust
mass for exceedingly elongated grains ($l/a > 3\times 10^4$) 
calculated from the antenna theory are much larger than those
from the infinite cylinder approximation (see Figure \ref{fig:cabs2v}). 
However, it is not clear whether the antenna theory applies to 
thin needles with a thickness of nanometer/micrometer in size. 
Unfortunately, the Discrete Dipole Approximation, the currently 
most powerful tool for solving the light scattering problem of 
non-spherical grains, is limited to grains of small size parameters
and small refractive indices (Draine 1988). 
Lack of an accurate solution, we are therefore 
not at a position to either approve or disprove the antenna theory 
for exceedingly thin needles.

\section{Conclusion}\label{sec:conclusion}
A wide variety of work have proposed the cosmic needle model as 
the CMB thermalizing agent: if cosmic metallic needle-shaped grains
absorb strongly at all wavelengths from IR to microwave wavelengths
it is possible to ascribe the observed background radiation at 
frequencies greater than $1\cm^{-1}$ as originating from thermalization, 
by these slender needles, of the radiation of Population III objects.
It is pointed out here that in many of these results insufficient attention
was given to the limits of applicability of the small particle
(Rayleigh) approximation. It is shown that the widely adopted Rayleigh
approximation is not applicable to conducting needles capable of 
supplying high far-IR and microwave opacities.
Due to lack of an accurate solution to the absorption properties of 
slender needles, we model them either in terms of infinite cylinders 
or the antenna theory. 
It is found that the available intergalactic iron dust,
if modelled as infinite cylinders, is not sufficient to produce 
a sufficiently large optical depth at long wavelengths required 
by the observed isotropy and Planckian nature of the CMB. 
In the context of the antenna theory, conducting needles with 
exceedingly large elongations ($>10^4$) appear to be capable of
satisfying the optical depth requirement without violating the
iron density limit. But the applicability of the antenna theory  
to exceedingly thin needles of nanometer/micrometer in thickness
needs to be justified.

\acknowledgments
I am grateful to A. Burrows, A.G.W. Cameron, B.T. Draine, J.I. Lunnine, 
..., and the late J.M. Greenberg for valuable discussions and
suggestions. I thank The University of Arizona for 
the ``Arizona Prize Postdoctoral Fellowship in 
Theoretical Astrophysics''.

\end{document}